\documentclass[a4paper,11pt]{article}
\usepackage{pos}

\usepackage{natbib}
\title{Implications of Mini-EUSO measurements for a space-based observation of UHECRs }

\author*[a] {Mario Bertaina}
\author[b] {Matteo Battisti}
\onbehalf{for the JEM-EUSO Collaboration\\[-1mm]{\normalsize \normalfont (a complete list of authors can be found at the end of the proceedings)}}
\affiliation[a]{Universit\`a di Torino \& INFN Sezione di Torino,\\}
\affiliation[b]{INFN sezione di Roma Tor Vergata,\\
}

\emailAdd{bertaina@to.infn.it}

\abstract{Mini--EUSO (Multiwavelength Imaging New Instrument for the Extreme Universe Space Observatory, known as
\emph{UV atmosphere} in the Russian Space Program) is the first mission of the JEM-EUSO program on board the
International Space Station. It was launched in August 2019 and it is operating since October 2019 being located in the Russian section (Zvezda module) of the station and viewing our planet from a nadir-facing UV-transparent window. The instrument is based on the concept of the original JEM-EUSO mission and consists of an optical system employing two Fresnel lenses of 25 cm each and a focal surface composed of 36 Multi-Anode Photomultiplier tubes, 64 channels each, for a total of 2304 channels with single photon counting sensitivity and an overall field of view of 44$^\circ \times $44$^\circ$. Mini-EUSO can map the night-time Earth in the near UV range (predominantly between 290 nm and 430 nm), with a spatial resolution of about 6~km and different temporal resolutions of 2.5~$\mu$s, 320~$\mu$s and 41 ms. Mini-EUSO observations are extremely important to better assess the potential of a space-based detector in studying Ultra-High Energy Cosmic Rays (UHECRs) such as K-EUSO and POEMMA. In this contribution we focus the attention on the results of the UV measurements and we place them in the context of UHECR observations from space, namely the estimation of exposure for the planned M-EUSO (Multi-messenger Extreme Universe Space Observatory) mission.}

\ConferenceLogo{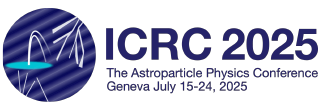}

\FullConference{39th International Cosmic Ray Conference (ICRC2025)\\
 15–24 July 2025\\
Geneva, Switzerland\\}

\begin{document}
\maketitle

\section{Introduction}
The current main goal in the field of UHECR (Ultra-High Energy Cosmic Ray) science is to identify their
astrophysical sources and acceleration mechanisms~\cite{snowmass}.
For this, increased statistics at the highest energies is one of the essential requirements. A space-based detector for
UHECR research has the advantage of a very large exposure and a uniform coverage of the celestial sphere.
The aim of the JEM-EUSO program~\cite{Zbigniew_ICRC2025} is to bring the study of UHECRs to space.
The principle of observation is based on the detection of UV light emitted by isotropic ﬂuorescence
of atmospheric nitrogen excited by Extensive Air Showers (EASs) in Earth’s atmosphere and
forward-beamed Cherenkov radiation reﬂected at the Earth’s surface or at dense cloud tops.

The JEM-EUSO program includes missions on ground (EUSO-TA~\cite{eusota}), on stratospheric
balloons (EUSO-Balloon~\cite{eusobal}, EUSO-SPB1~\cite{spb1}, EUSO-SPB2~\cite{spb2}, PBR~\cite{pbr}),
and from space (TUS~\cite{tus}, Mini-EUSO~\cite{minieuso}) employing fluorescence
detectors to demonstrate the feasibility of the UHECR observation from space and
prepare for the large size missions K-EUSO~\cite{keuso}, POEMMA~\cite{poemma} and the currently proposed M-EUSO mission~\cite{Zbigniew_ICRC2025} in response to the M8 ESA call of spring 2025. 

Mini-EUSO is the first detector of the JEM-EUSO program to observe the Earth from the International Space
Station (ISS) and to validate from there the observational principle of a space-based detector for UHECR
measurements. This means primarily demonstrating that a space-based observatory has a sufficiently high duty cycle, meant
as the fraction of time in which the atmospheric or anthropogenic light sources do not prevent the observation of a UHECR from space,
as well as the capability to detect short light transients which show similarities in terms of either the light intensity or
pulse duration with what is expected from an EAS cascading in the atmosphere. 


\section{The Mini-EUSO instrument and operation}
\label{sec:minieuso}
Mini--EUSO~\cite{minieuso}
 is a telescope operating in the near UV
range, predominantly between 290~nm and 430~nm, with a square focal surface corresponding to a field of view
(FoV) of $\sim$44$^{\circ}\times44^{\circ}$. Its spatial resolution at ground level is
$\sim$6$\times$6~km$^2$, slightly varying with the altitude of the ISS and the pointing direction of the 
pixel. The detector size is $37 \times 37 \times 62$~cm$^3$, mainly constrained by the size of the 
nadir-facing UV transparent window in the Russian Zvezda module. 
Until April 2025 more than 150 sessions have been successfully performed.
Currently, only the first 44 sessions are fully available for analysis. However, the remaining data up to April 2025 have been downloaded to ground and will be soon ready as well.

The optics are based on two 25~cm diameter Fresnel lenses in Polymethyl methacrylate (PMMA).
The Mini--EUSO focal surface, or Photo Detection Module (PDM), consists of a grid of 36 Multi-Anode
Photomultiplier Tubes (MAPMTs) 
arranged in an array of $6 \times 6$ elements.
Each MAPMT consists of $8 \times 8$ pixels, resulting in a total of 2304 channels.
The system collects data with three different time resolutions at the same time. The 2.5 $\mu$s GTU data stream (D1) is saved in a running buffer on which
the trigger code is executed (further details in \cite{matteo-trigger}). 
Similarly, an independent data stream with a dedicated trigger system and a time resolution of 320$\mu$s (D2) is present.
Finally, the longest time resolution of 40.96 ms (D3) is continuously stored.
An end-to-end in-flight calibration of the detector has been 
performed by assembling different UV-flasher systems on ground in Japan, Italy, and France 
and by firing them in various observational campaigns~\cite{calibration}. It made use of a 
flasher system consisting of a calibrated array of nine 100W COB-UV LEDs, batteries and an Arduino 
circuit. 
In 2022, an observational campaign was 
performed at Sant'Antimo Abbey (Italy) at the altitude of $\sim$320 m a.s.l. The 
place was chosen based on the very low light pollution in an area of several km radius. 
Photons from the UV flasher were detected and the overall 
efficiency of Mini-EUSO was estimated to be $\epsilon_{ME}$ = 0.061 $\pm$ 0.020 at
 $\lambda$ = 400 nm.\footnote{Typically, Mini-EUSO data are equalized through the use of a flat field map, to remove the differences in efficiency among the pixels. This value represents the efficiency of the flat fielded data.}

\section{Typical background variability during an orbit of Mini-EUSO}
\label{sec:orbit}
\begin{figure}[ht]
\centering
\includegraphics[width=\columnwidth]{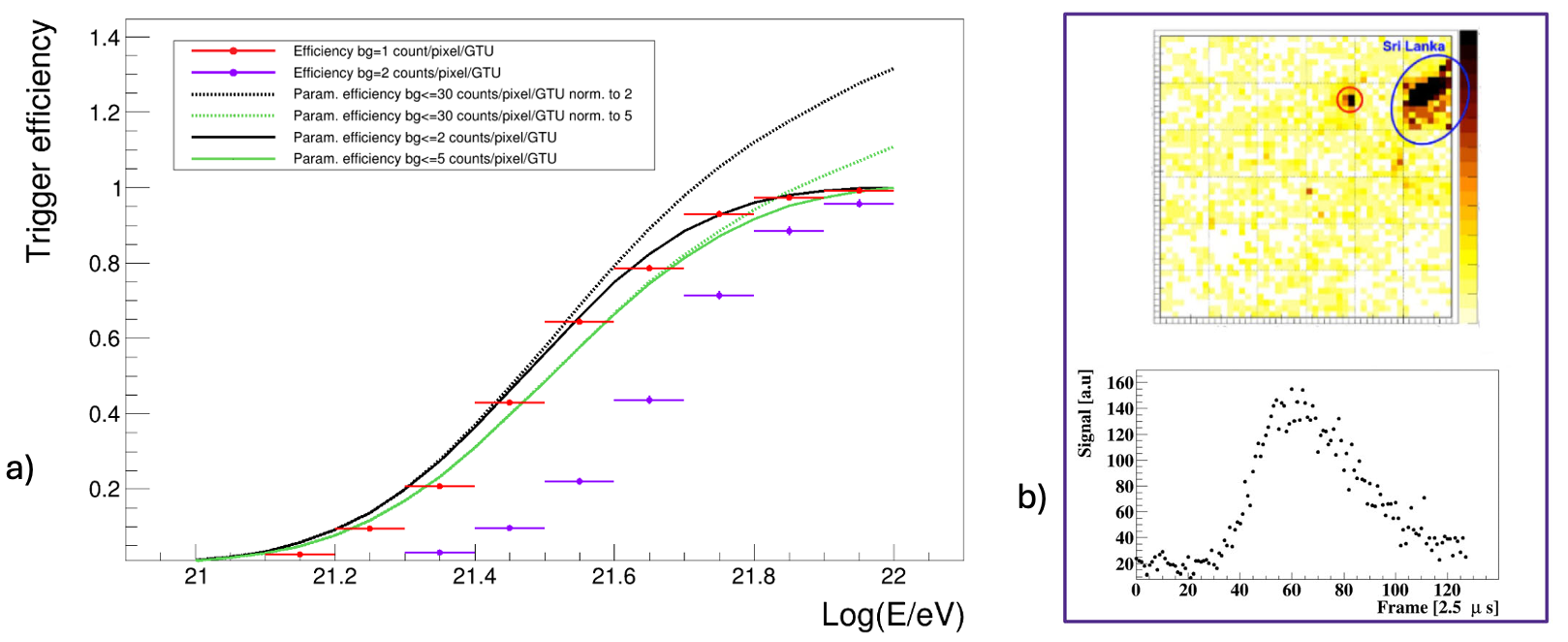}
\caption{a) The figure shows the trigger efficiency of Mini-EUSO as a function of energy for 
different background conditions (see Sect.~\ref{sec:exposure} for details).
b) Image (top) and light profile (down) of a Short Light Transients detected by
Mini-EUSO near Sri Lanka. The event has duration and light intensity similar to what is 
expected from an EAS signal even though their kinematics do not match (see discussion on Sect.~\ref{sec:exposure}).
}
\label{fig:expo-slt}       
\end{figure}

Mini-EUSO has been designed to detect a photon rate per pixel from diffuse sources (nightglow, clouds, cities, etc.) in the range of values expected from a large mission in space such 
as the original JEM-EUSO mission~\cite{jemeuso-exposure}. 
The pixel FoV is, therefore, $\sim$100 times larger in area with respect to the FoV of a 
JEM-EUSO pixel (0.5~km~$\times$~0.5~km), to compensate for the optical system $\sim$100 
times smaller, constrained by the dimension of the UV transparent window where Mini-EUSO is 
installed during the data taking sessions. As a consequence, the energy threshold of 
Mini-EUSO for UHECRs is well above 10$^{21}$~eV
(see Fig.~\ref{fig:expo-slt}).

\begin{figure}[h]
\centering
\includegraphics[width=\columnwidth]{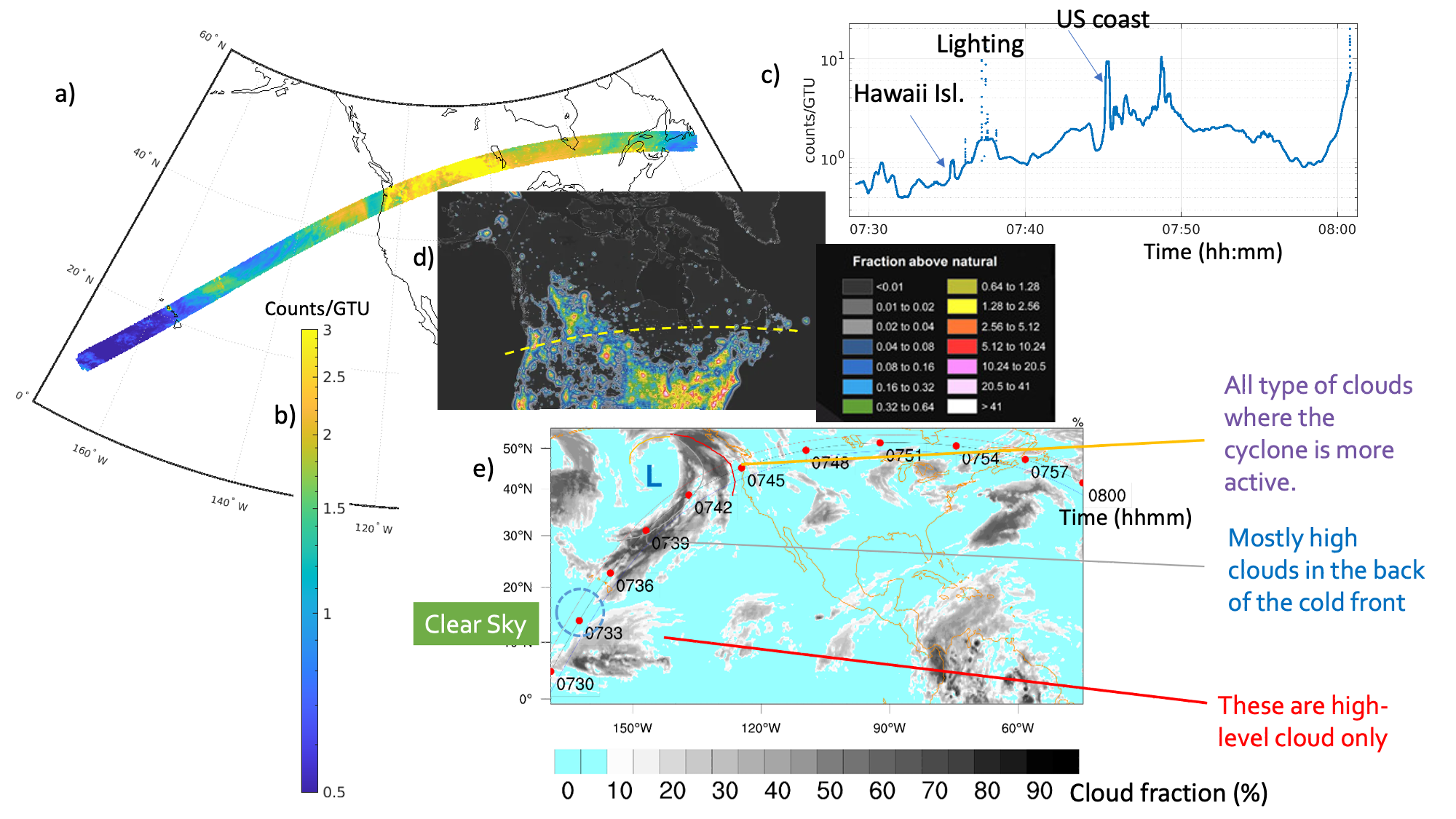}
\caption{The figure shows a portion of an orbit of Mini-EUSO on Session 30 (January 9$^{th}$ 
2021). Image a) shows the UV map of Mini-EUSO as a function of 
longitude and latitude. The counts/pix/GTU are color-coded in image b). They have been taken 
in D3 mode but normalized to the D1 GTU. Plot c) displays the light curve along the orbit of 
one pixel of Mini-EUSO. Plot d) presents the map of artificial lights taken from~\cite{falchi} 
(with a scale in fraction above natural light intensity), while e) the atmospheric conditions 
during the passage over ocean by means of a GFS post-processing analysis of the weather 
forecasts. The orbit of Mini-EUSO is highlighted by the three lines indicating the center and 
edges of the FoV. See text for details.
}
\label{fig:clouds-session30}       
\end{figure}

Fig.~\ref{fig:clouds-session30} shows a portion of Mini-EUSO orbit on Session 30, which took
 place on January 9$^{th}$ 2021 around 7:30 UTC. It displays an example of the different 
conditions that Mini-EUSO observes during an orbit: clear sky, cloudy conditions, city 
lights and land. Image a) shows the UV map as a function of longitude and latitude. The 
counts/pix/GTU are color coded in image b). They have been taken in D3 mode but normalized 
to the D1 GTU. Plot c) displays the light curve along the orbit of one pixel of Mini-EUSO. 
Plot d) presents the map of artificial lights in US and Canada~\cite{falchi} where the 
intensity is expressed in fractions above natural light intensity, while e) the atmospheric 
conditions during the passage over ocean by means of a post-processing analysis of the 
weather forecast obtained with the data collected by the Global Forecast System 
(GFS)~\cite{gfs}. The orbit of Mini-EUSO is 
highlighted by the three lines indicating the center and edges of the FoV of Mini-EUSO. 
As expected the lowest counts are detected over ocean on clear atmospheric conditions. In 
the presence of clouds the counts increase and the color contour matches in general the 
shape of the cloud location. The city lights of Hawaii islands are clearly visible in both 
a) (yellow spot) and c) (sharp peak increase of light). The presence of thunderstorm 
activity in the cloudy area between 7h36 and 7h39 UTC can be deduced by the pixel data 
reported in c) which show rapid increase and decrease of photon counts. 
Before entering the US coast there is a less cloudy region and the counts tend to decrease. 
Afterwards, when entering the US west coast sudden peaks in photon counts are observed. They
are due to the passage on particularly bright areas as depicted in map d). The counts 
remain higher than over the ocean also in the much less urbanized areas of Canada. This is 
due to the presence of snow which is highly reflective. The slightly higher counts on the  
St. Lawrence Bay in Quebec are due to the presence of low clouds. When Mini-EUSO goes back 
over the ocean the light intensity decreases again to ocean levels. The final sudden 
increase of counts is due to the sunrise which ends the night time of the orbit. 
This example shows the capability of the UV camera of Mini-EUSO to infer autonomously 
different atmospheric or ground scenarios encountered during the flight.
Mini-EUSO results on the average UV emissions in different conditions: clear and cloudy 
conditions, sea and land, various lunar phases are reported in~\cite{minieuso-uv}.
Assuming no-moon conditions and typical land/ocean and clear/cloudy atmosphere ratios
equal to 30/70, the average background level is $\sim$1.3 counts/pix/GTU, slightly below
1 counts/pix/GTU on ocean and clear sky, while a factor 1.5 - 2 higher in presence of clouds.
The average counts on land are $\sim$1.5 times higher than on ocean due to the tail of the 
distribution associated to the passage over civilized areas. However, deserts and forests
have much lower background values, ideal for collecting low energy events.

\section{Mini-EUSO duty cycle and exposure}
\label{sec:exposure}
\begin{figure}[hbtp]
\centering
\includegraphics[width=.95\textwidth]{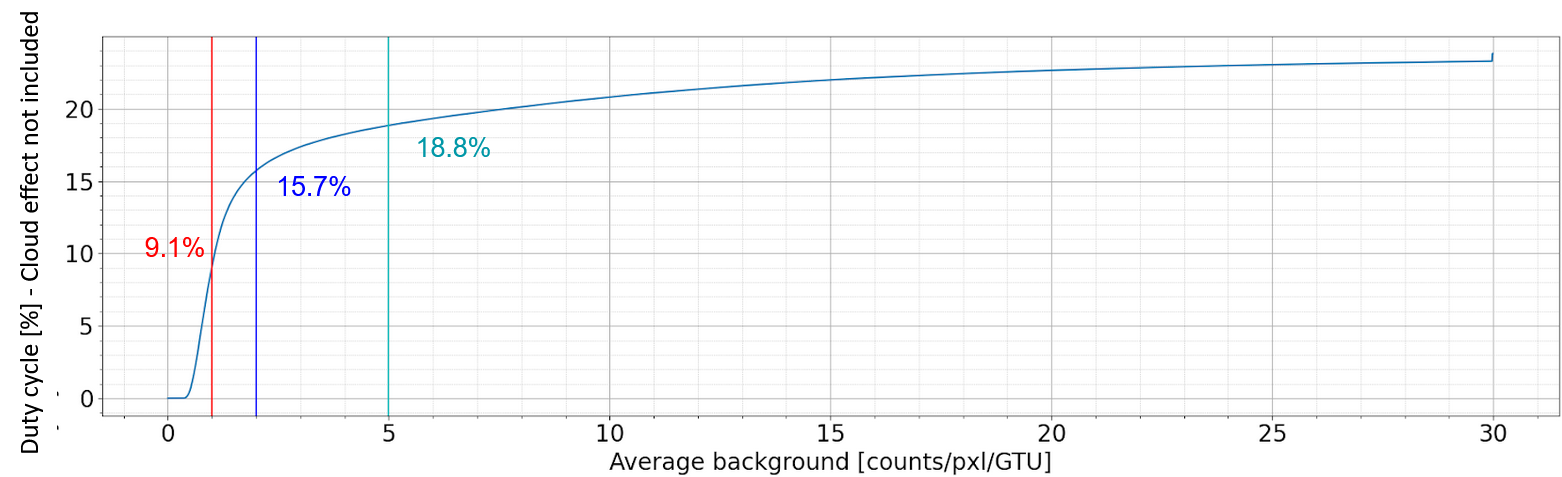}
\caption{Fraction of time (FoT) a space-based detector would spend below a given background level (BL), accounting for the diurnal and lunar cycles, as well as anthropogenic light sources. The values assume a Mini-EUSO detector efficiency of 8.7\%, consistent with that used in ESAF simulations. BLs below 0.4 cts/pix/GTU are never observed, while the last bin integrates the FoT in which it exceeded 30 cts/pix/GTU.}
\label{fig:Duty_cycle}
\end{figure}

Mini-EUSO data have been utilized to estimate the duty cycle as a function of background light for future space-based UHECR missions, such as the original JEM-EUSO, or the newly proposed M-EUSO mission. Using the complete Mini-EUSO dataset, we counted the number of instances in which each pixel recorded a given background level. Since Mini-EUSO preferably operates under favorable lunar conditions (i.e., near the new Moon, corresponding to lower background levels), we applied a correction to account for this observational bias.
Additionally, we evaluated whether the geographical distribution of Mini-EUSO’s observations - specifically the time spent over land versus ocean - is representative of a future space-based detector operating continuously. The resulting duty cycle, after applying all the aforementioned correction factors, is presented in Fig.~\ref{fig:Duty_cycle}.

As shown in the figure, approximately 9\% (16\%) [19\%] of the total observation time corresponds to background levels below 1 (2) [5] counts/pixel/GTU. These values incorporate corrections for the day/night cycle, Moon phase, and anthropogenic light sources, but do not account for cloud coverage. It is important to note that the 8.7\% detector efficiency assumed here aligns with the value used in ESAF simulations, which were also employed to generate the trigger efficiency curves shown in .\ref{fig:expo-slt}a). These curves represent the trigger performance of Mini-EUSO as a function of energy under varying background conditions, based on ESAF-simulated proton-induced EASs\cite{esaf}.
The curves are obtained either with fixed background conditions of 1 or 2 
counts/pix/GTU, or as cumulative distributions. In this second case the cumulative trigger 
efficiency curve is determined as a weighted sum of each trigger efficiency curve obtained for 
a specific background level and for the fraction of time in which it occurred. The weighted sum
 stops at 2 or 5 counts/pix/GTU or extends till 30 counts. The fraction of time in which the 
measured background by Mini-EUSO is less than 2(5) counts corresponds to a duty cycle of 
$\sim$16\%(19\%). 
The range of sessions between 20 and 44 have been used to perform this calculation which 
corresponds to the period in which Mini-EUSO operated in nominal conditions. 
The fact that both black and green curves tightly enclose the red points obtained at the fixed 
background level of 1 count/pix/GTU, which was assumed in the pre-flight simulations, 
confirms the derived duty cycle. 
This value is very similar to what was considered in the simulations of 
JEM-EUSO exposure~\cite{jemeuso-exposure}. The effect of clouds is not included yet. In JEM-EUSO a 
cloud efficiency of $k_c$ = 0.72 was estimated. More details on the methodology to derive 
Mini-EUSO exposure curves and to determine the duty cycle can be found in~\cite{uhecr2024}. 

The geometrical aperture of Mini-EUSO is obtained by multiplying the footprint of 
each pixel for 2304 available pixels. Considering the solid angle and the same cloud efficiency
 $k_c$ of JEM-EUSO, the collected exposure in these 25 sessions corresponds to 1200 Linsley at 
energies with full efficiency. The projected exposure considering all the data sessions from 
number 20 to 150 gives a value of $\sim$6500 Linsley. This value is comparable to the 
one accumulated by ground experiments with hybrid data.
Finally, an extrapolation, in the case of continuous operation of Mini-EUSO during the period October 2019 - April 2025 
years, leads to a potentially accumulated exposure at the highest energies of 
$\sim$165,000 Linsley, which is comparable with that collected so far by the surface arrays of Pierre Auger Observatory (PAO) and Telescope Array (TA). 
These results, despite being obtained with simplified assumptions, provide an idea of the 
potential of a space-based observatory devoted to the study of UHECRs.

The D1 trigger logic aims at detecting short pulses with time scales 
comparable to EASs. 
The most interesting
 categories of most frequent detected events are Ground Flashers (GFs) and Short Light Transients (SLTs). They are both 
identified by visually searching for short pulses which are either repeated or not. Sometimes 
they trigger two consecutive packets and the full shape is visible. Those with a duration 
$\leq200~\mu$s were looked at more carefully to check their compatibility with EAS-like events.
The visual inspection of the events identified 561 different GFs, some of which with comparable
light intensity or event duration of EASs. However, their periodic repetition and a careful 
comparison with EAS profiles and images on the FS allowed to discriminate them. The same 
inspection identified 14 SLTs (one of them is shown in Fig.~\ref{fig:expo-slt}b). The origin of
 those fast flashing lights is still under study, but it seems safe to assume that at least 
some of them are linked to thunderstorm activity. None of them could be 
misunderstood with an EAS~\cite{slt}.

\section{Preliminary expected performance of M-EUSO based on Mini-EUSO results}
\label{sec:meuso}
The JEM-EUSO collaboration respondend to the recent ESA call for M8 class missions by proposing the M-EUSO payload~\cite{Zbigniew_ICRC2025},
which consists of a single wide-field UV optical system coupled to two complementary detectors optimized for fluorescence and Cherenkov light, respectively. 
The key components are a refractive Fresnel lens assembly with 2.3~m diameter ($\gtrsim$4~m$^2$ aperture), providing a square-equivalent FoV of $\sim$40$^\circ$$\times$40$^\circ$ for the fluorescence camera. This corresponds to a pixel angular resolution of $0.12^\circ$. The 10~mm-thick lenses will be fabricated from UV-transparent PMMA. The fluorescence camera comprises 49 PDMs based on MAPMTs, with a total of $\sim$112,896 pixels offering single-photon sensitivity. It operates with 2.5~$\mu$s sampling for EAS reconstruction and longer integration times (320~$\mu$s and 40.96~ms) for complementary science objectives. The Cherenkov camera is mounted at the edge of the FoV and based on silicon photomultipliers (SiPMs), totaling $\sim$6000 pixels with comparable angular granularity. Operating at 10~ns sampling, it is optimized to detect fast, beamed Cherenkov light from near-horizontal showers directed toward the telescope. These include upward-going EASs from Earth-skimming $\nu_\tau$ interactions below the limb, as well as highly inclined UHECRs and neutrino-induced EASs skimming the atmosphere just above the limb. 
An additional compact infrared camera will monitor cloud coverage within and around the main FoV, ensuring accurate exposure determination. 

To ensure optimal sky coverage, M-EUSO will operate in low Earth orbit (LEO) at an altitude of approximately 450~km and an inclination of $\sim$30$^\circ$. This configuration guarantees near-uniform full-sky exposure over the course of the mission, while avoiding the exposure deficits at high declination that affect lower-inclination orbits.
M-EUSO will operate in two complementary observational modes to meet its full science goals across the ultra-high-energy regime: a nadir mode for UHECR studies at the lowest possible threshold, and a tilted mode optimized for neutrino detection and for enhancing exposure at the highest energies.

Based on the Mini-EUSO-derived duty cycle the expected annual exposure in nadir mode is $\sim$30,000~km$^2$~sr~yr. This exceeds, in just one year, the total lifetime exposure of TA, and corresponds to 25\% of the exposure accumulated by the PAO over 20 years. 
After this initial phase, M-EUSO will move to tilted mode, pointing toward the Earth’s limb. Although the energy threshold increases due to longer light paths and lower photon collection efficiency, the accessible atmospheric volume and geometrical aperture grow substantially. In this configuration, the mission will reach $\sim$100,000~km$^2$~sr~yr per year above $3 \times 10^{20}$~eV. 


M-EUSO’s expected performance is based on consolidated simulations using the ESAF framework~\cite{esaf}, developed and benchmarked through the design studies of K-EUSO and POEMMA. The fluorescence camera parameters are grounded in actual calibration data from EUSO-SPB2, while background and duty cycle assumptions are derived from in-orbit measurements by Mini-EUSO. 
M-EUSO is expected to have an energy resolution of $\sim$30(20)$\%$ at $\sim$5$\times 10^{19}$(3$\times 10^{20}$) eV and an angular one of $\sim$5$^{\circ}$($2^{\circ}$) for 60$^\circ$ inclined EAS.
These capabilities support robust reconstruction of the UHECR energy spectrum and sky distribution, enabling anisotropy studies and comparisons with nearby extragalactic structure catalogs. The angular resolution is sufficient to search for source correlations above the expected magnetic deflection scale at the highest energies.


\section{Acknowledgements} 
This work was supported by the Italian Space Agency (ASI) through the agreement between ASI and University of Roma Tor Vergata 2020-26-HH.0, by the French space agency CNES, and by the National Science centre in Poland grant 2020/37/B/ST9/01821. This research has been supported by the Russian State Space Corporation Roscosmos.


\bibliography{my-bib-database}



    \newpage
{\Large\bf Full Authors list: The JEM-EUSO Collaboration}

\begin{sloppypar}
{\small \noindent
M.~Abdullahi$^{ep,er}$              
M.~Abrate$^{ek,el}$,                
J.H.~Adams Jr.$^{ld}$,              
D.~Allard$^{cb}$,                   
P.~Alldredge$^{ld}$,                
R.~Aloisio$^{ep,er}$,               
R.~Ammendola$^{ei}$,                
A.~Anastasio$^{ef}$,                
L.~Anchordoqui$^{le}$,              
V.~Andreoli$^{ek,el}$
% ...
}
\end{sloppypar}
\vspace*{.3cm}

{ \footnotesize
\noindent
%
% ...
$^{cb}$ Universit\'e de Paris, CNRS, AstroParticule et Cosmologie, F-75013 Paris, France\\
% ...
$^{ef}$ Istituto Nazionale di Fisica Nucleare - Sezione di Napoli, Italy\\
% ...
$^{ei}$ Istituto Nazionale di Fisica Nucleare - Sezione di Roma Tor Vergata, Italy\\
% ...
$^{ek}$ Istituto Nazionale di Fisica Nucleare - Sezione di Torino, Italy\\
$^{el}$ Dipartimento di Fisica, Universit\`a di Torino, Italy\\
% ...
$^{ep}$ Gran Sasso Science Institute, L'Aquila, Italy\\
$^{er}$ INFN, Gran Sasso National Laboratory, Assergi (AQ), I-67100, Italy\\
% ...
$^{ld}$ University of Alabama in Huntsville, Huntsville, AL, USA\\
$^{le}$ Lehman College, City University of New York (CUNY), NY, USA\\
% ...
}

\end{document}